\documentclass[aps,prl,twocolumn,footinbib,showpacs]{revtex4}
\usepackage{amsmath,amssymb}
\usepackage{graphics,color}

\renewcommand{\Re}{\textrm{Re}}
\renewcommand{\Im}{\textrm{Im}}

\begin{document}

\title{Localized Modes in a Finite-Size Open Disordered Microwave Cavity}
\author{David Laurent, Olivier Legrand, Patrick Sebbah, Christian Vanneste, Fabrice Mortessagne}
\email{fabrice.mortessagne@unice.fr}
\affiliation{Laboratoire de Physique de la Mati\`ere Condens\'ee,
CNRS UMR 6622,\\  Universit\'e de Nice Sophia-Antipolis, 06108 Nice,
France}

\begin{abstract}
We present measurements of the spatial intensity distribution of localized modes in a two-dimensional open microwave cavity randomly filled with cylindrical dielectric scatterers. We show that each of these modes displays a range of localization lengths and successfully relate the largest value to the measured leakage rate at the boundary. These results constitute unambiguous signatures of the existence of strongly localized electromagnetic modes in two-dimensionnal open random media.
\end{abstract}

\vskip.6cm

\maketitle
In spite of two decades of intensive research, the experimental observation of strong localization of classical waves remains a tremendous challenge \cite{GenackChabanov2005}. One of the main difficulties lies in the fact that the signature of localization is sought after through statistic measurements of transmission. Indeed, the expected exponential decrease of transmission could not be attributed unambiguously to localization rather than absorption  \cite{Wiersma97}. In contrast, transient analysis, as demonstrated by Weaver in ultrasound measurements in a two-dimensional (2D) disordered medium \cite{Richard}, allows to discriminate between these two phenomena. A similar approach has been proposed in disordered microwave systems by Genack who demonstrated localization in the microwave regime using variance of total transmission as a genuine probe to localization even in the presence of strong absorption \cite{Chabanov2000}. More recently, G. Maret's group have observed deviation from diffusion in the time of flight distribution, which cannot be explained by absorption but plead for a transition to strong localization of photons in three dimensions \cite{Storzer2006}. Such signatures of localization in transmission reflects the nature of the underlying quasimodes, which are spatially localized inside the system. The increasing contribution of such long-lived modes, as time progresses, is responsible for the observed deviations from the regime of  purely diffusive transport \cite{Chabanov_PRL_2003}. Besides the fact that their localized nature is not affected by absorption, exhibiting the localized modes inside the random system would be a direct demonstration of localization and the key to the understanding of the mechanism underlying the transition from a diffusive to a localized regime.

Observations of localized modes have been mostly reported in 1D random systems \cite{1D} where localization is readily achieved. In the marginal dimension of 2, localization of bending waves in randomly loaded steel plates \cite{Ping} and spatial concentration of microwave field in microwave cavities \cite{Dalichaouch91} have been reported where the influence of the reflecting boundaries precludes the unambiguous attribution of the localization effect to the sole presence of scatterers. The aim of this letter is to provide an unequivocal observation of localized modes in an open 2D random system. Beyond the mere observation of such modes, we put them under a test of robustness and successfully confront them to numerical simulations and theoretical predictions concerning the relationship between their spectral widths and the localization lengths.

\begin{figure}[h]
\begin{center}
\includegraphics{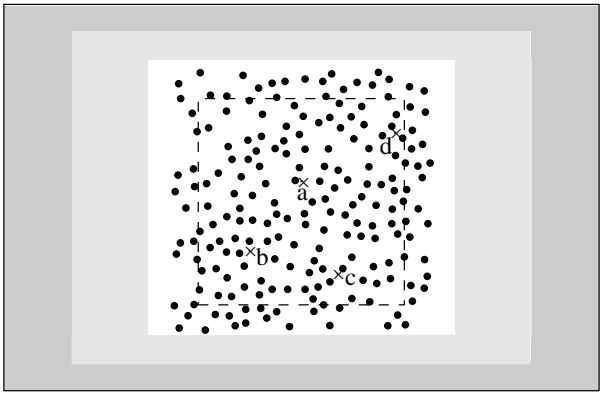}
\end{center}
\caption{Schematic view of the cavity: 196 scatterers located in a central square region. For some measurements, the scatterers outside the dashed square are withdrawn. Two different layers of microwave absorbers ensure a weak inward-reflection of the field. The antennas (a, b, c, d) are shown through small crosses.}
\label{cavity}
\end{figure}

 All our results are obtained from transmission signals measured in a 2D microwave cavity operated at frequencies ranging from 500\,MHz to 10\,GHz. The original rectangular ($76\times47$\,cm) bare cavity is composed of two copper plates sandwiching a copper rectangular frame of thickness $5$\,mm.  The quality of copper is OFHC (Oxygen-Free-High-Conductivity) to reduce ohmic losses.  Due to its height of 5\,mm (smaller than half the smallest wavelength used), this cavity only admits transverse magnetic two-dimensional modes of order 0. Through one of the copper plates, a few antennas are introduced, their optimal weak coupling being obtained by fixing their penetration length inside the cavity at 2\,mm. The antennas are monopolar with SMA connectors and their positions are displayed on Fig. \ref{cavity}. In transmission measurements, only two antennas are used at a time, the other ones being terminated by 50\,$\Omega$ loads so that all antennas behave the same way regarding the losses they imply. These antennas are linked to an HP\,8720\,D vector analyzer through flexible cables. The details about transmission measurements are described in reference  \cite{Jerome_2004}. 

In order to mimic a truly open system, frames of microwave absorbing foams are inserted in the cavity, surrounding an empty rectangular-shaped space. Two different layers of absorbers (ECCOSORB$^\circledR$ LS-14 and LS-16) are used with different electromagnetic impedances (see Fig. \ref{cavity}), the inner layer having a relative impedance of $0.89$ and the outer $0.87$ at 10\,GHz. Their thickness is the overall thickness of the cavity and their respective widths are determined to provide 20\,dB of attenuation in the intensity reflected back to the empty space at frequencies above 500\,MHz. As shown in Fig. \ref{cavity}, a disordered medium is introduced in the empty space. It consists of 196 cylindrical dielectric scatterers with a measured dielectric constant $\epsilon =37$, a radius of 3\,mm and a height of 5\,mm, the scattering region being a 25\,cm $\times$ 25\,cm square. The scatterers have been chosen for their particular low loss (quality factor $Q=7000$ at 7\,GHz) and their high index of refraction. In the frequency range of interest, the possible excitation of transverse electric waves inside the dielectric scatterers can be assumed to be negligible. It is worth noting that our 2D system is formally equivalent to the problem of a quantum particle with positive energy above a disordered potential consisting of negative circular wells embedded in a zero potential. To be more explicit, at a given frequency $\omega = c k$, the 2D Helmholtz equation for the component $\phi$ of the electric field perpendicular to the plane of the cavity, can be written in the form of the stationary Schr\"odinger equation
\begin{equation}
\label{eigprob}
\left[-\Delta
+(1-n^2(\vec{r}))k^2\right]\,\phi(\vec{r})
=k^2\,\phi(\vec{r})\,.
\end{equation}
Therefore, the well associated to a dielectric scatterer should not be viewed as a confining well but rather as a resonant scattering well.

\begin{figure}[h]
\begin{center}
\includegraphics{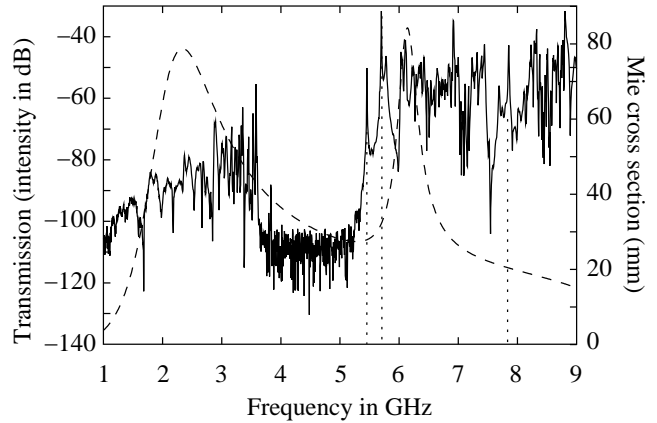}
\end{center}
\caption{Full curve : transmitted intensity between antennas a and b for the disordered realization shown in Fig. \ref{cavity}. Dashed curve: Mie cross section of a single dielectric scatterer. The dotted lines indicate the resonances whose associated wavefunctions are depicted on Fig. \ref{cartosim}.}
\label{spectrum}
\end{figure}

The positions of the scatterers are determined at random with a filling fraction of 8.9\,\% and a minimal distance $d_\textrm{min}$ between the centers of neighboring scatterers. At the above given filling fraction, the values of $d_\textrm{min}$ range from 8\,mm to 19\,mm. The smallest value is dictated by practical constraint, the largest corresponds to the lattice parameter of a square array. We fixed $d_\textrm{min}=11$\,mm. Figure \ref{cavity} displays a particular realization of disorder using the above parameters. In Fig. \ref{spectrum}, a typical transmission spectrum (between antennas a and b) is shown. One can clearly identify a frequency range which is reminiscent of the bandgap generally found in the case of a periodic structure. Here the central frequency of the observed gap is essentially related to the neighboring Mie resonances \cite{Sigalas96,Ping} (the Mie cross section of a single scatterer is shown superimposed to the measured spectrum in Fig. \ref{spectrum}). The width of the gap is controlled by $d_\textrm{min}$. Using a \emph{diffusion based theory} \cite{Derode2001,VandeHulst, Gupta2003} we evaluated the localization length $\xi_{theory}$ in two dimensions  for our system of scatterers within the independent scattering approximation. It reads :
\begin{equation}
\xi_{theory} = \ell \exp ( \pi \Re (k_{eff}) \ell / 2)
\end{equation}
where $k_{eff}$ is the effective complex wave number and $\ell = \left( 2 \Im ( k_{eff} ) \right)^{-1}$ is the mean free path, which is inversely proportional to the Mie cross section. The minimal values of this localization length are roughly located at the maxima of the Mie cross section and are of the order of 10\,mm, a value comparable to the mean free path and also to the mean distance between scatterers for the filling fraction mentioned above.

By using an appropriate fitting procedure \cite{Jerome_2004}  we extracted the central frequency and the spectral width of a given resonance. Due to the finite size of the scattering region, only a few widths take values close to those expected when taking only the ohmic losses into account (close to 1\,MHz at 6\,GHz), whereas the vast majority are significantly larger. To begin, we focused our attention on three resonances: one in the \emph{gap} at 5.45\,GHz, one on the edge of the \emph{gap} at 5.66\,GHz, and one at 7.80\,GHz, and with respective widths 1.1\,MHz, 3.0\,MHz, 15.0\,MHz. When using different pairs of antennas we obtained spectra very similar to the one presented in Fig. \ref{spectrum}, the most affected resonances being the sharpest, like the three selected, as they can even disappear for a given pair. This behavior pleads in favor of the observation of strongly localized modes. Indeed, as will be shown below, when both antennas are located on vanishing tails of a given localized mode, the transmission signal itself vanishes.

We obtained the spatial distribution of these resonances by using a well-known scanning perturbation technique  (see ref. \cite{Gokirmak_1998} and references therein). This technique is based on the fact that, for a given resonance, a frequency shift is caused by the presence of a small movable stainless steel bead placed inside the cavity. This perturbation technique crucially relies on the fact that the analyzed resonance is sufficiently well isolated from its neighbors. The shift is a measure of the intensity of the local field at the bead location, and is essentially proportional to the square of the electric field\cite{Gokirmak_1998}. Indeed, through measurements on computable modes of an empty rectangular cavity we checked that the transverse magnetic field contribution is at most of the order of 1\,\% of the maximum contribution of the electric field. The bead can be moved on a very finely defined grid (5-mm-step) from the outside of the cavity by means of a strong magnet which is fixed on an X-Y precise translation stage.

\begin{figure}[h]
\begin{center}
\includegraphics{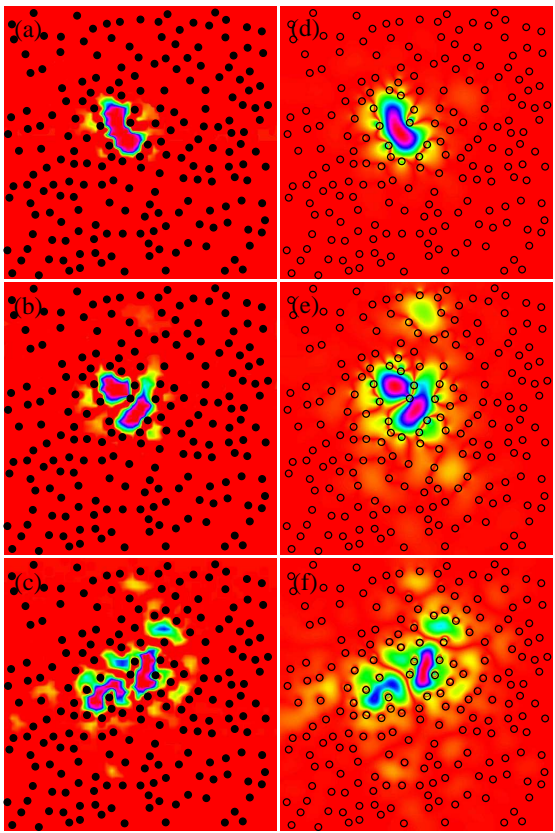}
\end{center}
\caption{(Color online) Comparison between experimental and numerical maps of the amplitude of the electric field for localized modes at 5.45\,GHz (a) and (d),  5.66\,GHz (b) and (e), and at 7.80\,GHz (c) and (f), the scattering region being a 25\,cm $\times$ 25\,cm square. The numerical maps are obtained through FDTD-based simulations (with perfectly open boundary conditions).}
\label{cartosim}
\end{figure}
In figure \ref{cartosim} (a,b,c), two-dimensional scans of the amplitude (square root of the intensity) of the three long-lived modes mentioned above are depicted. They display a clear spatial concentration. Whereas mode (a) seems to be trapped in an small region surrounded by scatterers, modes (b) and (c) are more extended. Following a method indicated below, we  evaluated the localization lengths for these modes: between 15 and 25\,mm for mode (a), 25-36\,mm for mode (b) and 38-50\,mm for mode (c) (the ranges depending on the anisotropy of the spatial pattern). Note that modes (a) and (b) have vanishing amplitudes at the locations of antennas b and c (see Fig. \ref{cavity}) and that their corresponding peaks vanish in the transmission signal when the pair of antennas b and c is used.

For the sake of completeness, we also performed numerical simulations of the Maxwell equations in our system through a Finite-Difference Time-Domain (FDTD) method \cite{Taflove}. To mimic perfectly open boundary conditions Perfectly Matched Layer (PML) boundaries were implemented \cite{Berenger95}. These PML conditions are used in the same spirit as the absorbing microwave layers of the actual experiment. In the numerical simulations, losses are only due to the leakage through the boundary since no ohmic dissipation is included. Hence, spectral responses display sharper peaks than in the experiment but all the identified resonances of the experiment are found at the corresponding frequencies. Fig. \ref{cartosim} shows a comparison between experimental and simulated modes. The agreement is good. Note the presence of a nonvanishing field inside the scatterers, such information being beyond the reach of our experimental measuring technique.

\begin{figure}[h]
\begin{center}
\begin{picture}(0,0)%
\includegraphics{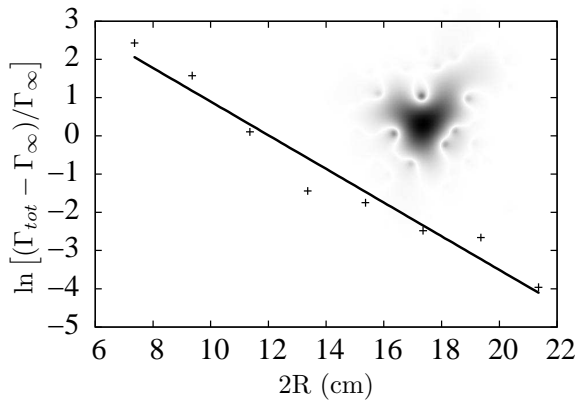}%
\end{picture}%
\setlength{\unitlength}{4144sp}%
\begingroup\makeatletter\ifx\SetFigFont\undefined%
\gdef\SetFigFont#1#2#3#4#5{%
  \reset@font\fontsize{#1}{#2pt}%
  \fontfamily{#3}\fontseries{#4}\fontshape{#5}%
  \selectfont}%
\fi\endgroup%
\begin{picture}(3386,2557)(1757,-3183)
\put(1889,-1841){\rotatebox{90.0}{\makebox(0,0)[b]{\smash{{\SetFigFont{10}{12.0}{\familydefault}{\mddefault}{\updefault}{\color[rgb]{0,0,0}$\ln\big[(\Gamma_{tot}-\Gamma_\infty)/\Gamma_\infty\big]$}%
}}}}}
\put(3646,-3143){\makebox(0,0)[b]{\smash{{\SetFigFont{10}{12.0}{\familydefault}{\mddefault}{\updefault}{\color[rgb]{0,0,0}2R (cm)}%
}}}}
\end{picture}%
\end{center}
\caption{Evolution of the reduced normalized width vs the size $2R$ of the scattering system for the mode shown in inset. $\Gamma_{\infty}$ is the contribution of ohmic losses to the total spectral width $\Gamma_{tot} = \Gamma_{leak} + \Gamma_{\infty}$. Straight line is the best linear fit whose absolute value of the inverse slope is $2.3\pm 0.2\,$cm}
\label{GammavsR}
\end{figure}

As the numerical modes have precisely calculated exponentially vanishing tails far from their centers, we extracted values of their localization lengths $\xi_{loc}$ by numerically evaluating the field amplitude spatial autocorrelation. (Due to a small contribution from the squared magnetic field to the frequency shift, our experimental scanning technique is less accurate for this purpose.) The first important result we deduce from the field amplitude spatial autocorrelation concerns the anisotropy of the exponential decay for a given mode. In our experimental system, the scale separation between the localization length and the size of the scattering system is most likely not sufficient to observe a spatial decay with a single characteristic length at large distances.  The overall values of $\xi_{loc}$ thus obtained are in complete agreement with the minimal values of $\xi_{theory}$. The finite size of the scattering system is also known to influence the spectral widths due to leakage at the boundary \cite{Pinheiro2004}. This influence is quantitatively described by the following relation between the spectral partial width $\Gamma_{leak}$ due to leakage at the boundary,
the localization length $\xi_{loc}$, and the distance $R$ from the center of the mode to the boundary:
\begin{equation}
\Gamma_{leak} \propto \exp (-2R/\xi_{loc})
\label{leak}
\end{equation}
This is readily demonstrated in our experiment. By reducing the size of the scattering region by steps of 2\,cm we experimentally obtained increasing values of the spectral width for a given resonance. We then extracted the partial width $\Gamma_{leak}$ from the measured total width $\Gamma_{tot}$ thanks to a proper evaluation of the contribution $\Gamma_{\infty}$ of Ohmic losses \cite{Jerome_2004}. The behavior of $\Gamma_{leak}$ given by relation (\ref{leak}) was observed, in different realizations of disorder, for many localized modes  provided they were sufficiently well centered in the scattering region (otherwise, the above reduction was not feasible). This is illustrated in Fig. \ref{GammavsR} for a mode measured at 4.2\,GHz in a scattering region of the same size as before with a filling fraction of 5.5\,\% and $d_{\textrm{min}}=13$\,mm. In Fig. \ref{GammavsR}, $\ln[(\Gamma_{tot}- \Gamma_{\infty})/ \Gamma_{\infty}]$ is plotted  vs the size $2R$ of the scattering region. The expected exponential behavior is quantitatively verified if one uses, instead of the localization length, a value close to the maximal value of the range deduced from the spatial autocorrelation: $1.4\,$cm $\le \xi_{loc}\le 2.5\,$cm. All this confirms that the exponentially vanishing field of such modes at the boundary involves all the scatterers of the system through complex multiple interference effects. In consequence, we can conclude that the modes we observe should not be simply considered as \emph{defect-like} modes or \emph{cavity} modes built by the mere presence of a few scatterers surrounding them. 
 
In summary, we have presented the unambiguous observation of strongly localized modes in a two-dimensional open disordered microwave cavity. Most importantly, we have investigated the influence of the finite size of the scattering system on the structure of localized modes. Indeed, we have found that the naive picture of exponentially decaying spatial envelopes should be reconsidered in view of the complex patterns of the observed modes. Nonetheless, through the field amplitude spatial autocorrelation we extracted a range of localization lengths for a given mode and successfully related the largest value to the measured leakage rate at the boundary.

\begin{acknowledgments}
We are very grateful to Gregory Sauder for the automation of the data acquisition. D. Laurent acknowledges financial support from DGA/CNRS Grant n. 2004487. This work is supported by the Groupement de Recherche IMCODE (CNRS n. 2253).
\end{acknowledgments}

\end{document}